\documentclass[intlimits,twoside,a4paper]{article}

\usepackage{amsmath,amssymb}
\usepackage{graphicx}

\usepackage[T2A]{fontenc}
\usepackage[cp1251]{inputenc}

\usepackage[eqsecnum]{cmpj2}

\issue{2013}{16}{4}{43803}
\doinumber{10.5488/CMP.16.43803}

\title[Madden-Glandt approximation for partly-quenched systems]
{Structure and thermodynamics of the primitive model electrolyte in a charged matrix: The evaluation of the Madden-Glandt approximation\thanks{Dedicated to Professor Myroslav Holovko on the occasion of his 70$^\mathrm{th}$ birthday.}}
\author[M. Luk\v{s}i\v{c}, B. Hribar-Lee]{M. Luk\v{s}i\v{c}\refaddr{label1,label2}, B. Hribar-Lee\refaddr{label2}}
\addresses{
\addr{label1} University of Ljubljana, Faculty of Chemistry and Chemical Technology,
\\ A\v{s}ker\v{c}eva c. 5, SI-1000 Ljubljana, Slovenia
\addr{label2} The State University of New York at Stony Brook,
Laufer Center for Physical and Quantitative Biology, \\
5252 SUNY, Stony Brook, NY 11794-5252, United States}

\date{Received August 2, 2013, in final form August 31, 2013}
\authorcopyright{M. Luk\v{s}i\v{c}, B. Hribar-Lee, 2013}

\begin{document}

\maketitle

\begin{abstract}
We compared the results of the Madden-Glandt (MG) integral equation approximation for partly-quenched systems with the commonly accepted formalism of Given and Stell (GS). The system studied  was a $+1:-1$ restricted primitive model (RPM) electrolyte confined in a quenched $+1:-1$ RPM matrix. A renormalization scheme was proposed for a set of MG replica Ornstein-Zernike equations. Long-ranged direct and total correlation functions, describing the interactions between the annealed electrolyte species within the same replicas and between the annealed and matrix particles, appeared to be the same for MG and GS approach. Both versions of the theory give very similar results for the structure and thermodynamics of an annealed subsystem. Differences between excess internal energy, excess chemical potential, and isothermal compressibility become pronounced only at high concentrations of matrix particles.
\pacs 82.45.Gj, 02.30.Rz, 61.20.-p, 82.60.-s
\end{abstract}

\section{Introduction}

In the last decades, much attention has been paid to the properties of electrolyte solutions adsorbed in random matrices (for review see reference \cite{pqs11}). These systems can be considered as partly-quenched, meaning that some degrees of freedom are quenched and others are annealed \cite{pqs11}. Such systems differ from regular mixtures: here all the statistical-mechanical averages needed to calculate the systems' thermodynamic and dynamic properties become double ensemble averages and the thermodynamic relations need to be rewritten accordingly \cite{given1992,given1994}.

In general, one can describe partly-quenched systems as consisting of two components (subsystems), one being quenched (usually called the \textit{matrix}) and the other being allowed to equilibrate in the presence of the matrix subsystem (annealed subsystem). Usually, it is assumed that the annealed fluid does not affect the matrix \cite{pizio1998,rosinberg1999,pizio2000,trokhymchuk1996}.

To develop the integral equation theory for partly-quenched systems, Madden and Glandt \cite{madden1988,madden1992} split the total correlation functions, $h(r)$, and direct correlation functions, $c(r)$, into a ``connected'' part, representing the interactions between annealed particles within the same replica, and a ``blocking'' part (here denoted by index 12), describing the interactions between the annealed particles mediated by the matrix particles. In their approximation, they set a blocking part of the direct correlation function to zero ($c^{12}(r) = 0$) and obtained a set of Madden-Glandt Ornsten-Zernike (MGOZ) equations \cite{madden1988,madden1992}
\begin{equation}
\begin{split}
{\bf h}^{00}-{\bf c}^{00} &= {\bf c}^{00}\otimes {\mbox{\boldmath ${ \rho}$}}^0{\bf h}^{00},\\[5pt]
{\bf h}^{10}-{\bf c}^{10} &= {\bf c}^{10}\otimes {\mbox{\boldmath ${ \rho}$}}^0{\bf h}^{00}+{\bf c}^{11}\otimes {\mbox{\boldmath ${ \rho}$}}^1{\bf h}^{10},
\\[5pt]
{\bf h}^{11}-{\bf c}^{11} &= {\bf c}^{10}\otimes {\mbox{\boldmath ${ \rho}$}}^0{\bf h}^{01}+{\bf c}^{11}\otimes {\mbox{\boldmath ${ \rho}$}}^1{\bf h}^{11}.
\end{split}
\label{mgoz}
\end{equation}
Index 0 denotes the matrix subsystem, index 1 is the annealed fluid subsystem, ${\mbox{\boldmath ${ \rho}$}}^i$ is the number density of the species $i$, and symbol $\otimes$ denotes convolution in $r$-space. The reasons for choosing  matrix notation will be given is section~\ref{sec:theor}.

Later on, Given and Stell derived the exact integral equations for such systems \cite{given1992,given1994}. They are commonly known as the Replica Ornstein-Zernike (ROZ) equations and read
\begin{equation}
\begin{split}
{\bf h}^{00}-{\bf c}^{00}&={\bf c}^{00}\otimes {\boldsymbol \rho^0}
{\bf h}^{00}, \\[5pt]
{\bf h}^{10}-{\bf c}^{10}&={\bf c}^{10}\otimes {\boldsymbol \rho^0}
{\bf h}^{00}+{\bf c}^{11}\otimes {\boldsymbol \rho^1} {\bf h}^{10}-
{\bf c}^{12}\otimes {\boldsymbol \rho^1}{\bf h}^{10},  \\[5pt]
{\bf h}^{11}-{\bf c}^{11}&={\bf c}^{10}\otimes {\boldsymbol \rho^0}
{\bf h}^{01}+{\bf c}^{11}\otimes {\boldsymbol \rho^1}{\bf h}^{11}-
{\bf c}^{12}\otimes {\boldsymbol \rho^1}{\bf h}^{21},  \\[5pt]
{\bf h}^{12}-{\bf c}^{12}&={\bf c}^{10}\otimes {\boldsymbol \rho^0}
{\bf h}^{01}+{\bf c}^{11}\otimes {\boldsymbol \rho^1}{\bf h}^{12} +
{\bf c}^{12}\otimes {\boldsymbol \rho^1}{\bf h}^{11}-
2{\bf c}^{12}\otimes {\boldsymbol \rho^1}{\bf h}^{21}.
\end{split} \label{roz}
\end{equation}
Here, ${\bf c}^{12}$ and ${\bf h}^{12}$ (${\bf h}^{21}$) stand for blocking parts of the direct and total correlation functions, respectively.

While both approaches have been used to study partly-quenched systems (for review see references \cite{pizio1998,rosinberg1999,pizio2000,pqs11} and references therein), only the ROZ equations were applied to systems where both subsystems (i.e., quenched and annealed) were electrolytes. One of the main reasons for that lies in the fact that due to the long-ranged nature of the Coulomb interaction, an appropriate renormalization scheme is necessary to solve the above equations within the integral equation theory framework \cite{ichiye1990,vlachy1991,duh1992,hribar1997,hribar1998}. To our best knowledge, there exist no documented attempt for the renormalization of MGOZ equations.

Therefore, in the present contribution we propose a renormalization scheme for the MGOZ equations and calculate the structural and thermodynamic properties of simple model systems: the quenched and annealed subsystems are both $+1:-1$ restricted primitive model electrolytes. We compare the results of MGOZ approach to the ROZ one. Hypernetted-chain (HNC) approximation was used as the closure relation in both cases \cite{hribar1998}. The comparison of the results for the Madden-Glandt and Given-Stell approach enables one to gain insights into the meaning of the ``blocking'' function. MG approximation seems also to be an easier starting point in developing theories of association and other related phenomena (see, for example, reference \cite{trokhymchuk1996}). However, there are reasons beyond academic ones why one should explore the MG approximation in the case of electrolyte solutions. One of them is the fact that the MG approach is in principle less demanding than the ROZ because the equations and the closure relations are simpler (contain less terms). The computationally convergence is, therefore, faster compared to ROZ.

We structured the paper as follows:  After an Introduction, a short description of the model is given, followed by the description of the renormalization procedure. Next, the results for a few typical cases studied are presented and conclusions are given in the end.

\section{The model}

The model studied here is basically the same as the one studied previously by the Monte Carlo simulation methods \cite{bratko1996}, as well as by the integral equation theory \cite{hribar1998}: the so-called primitive model for electrolyte solutions, where the particle-particle interaction $U_{ij}^{mn}(r)$ is given by
\begin{equation}
U_{ij}^{00}(r) = \left\{
\begin{array}{ll}
\infty\,, & \hbox{$r <  ( \sigma_i^0 + \sigma_j^0)/2$}\,, \\[5pt]
\displaystyle \frac{e^2z_i^0z_j^0}{4\pi \varepsilon_\mathrm{v}
 \varepsilon_0 r}\,, &
\hbox{$r\geqslant(\sigma_i^0 + \sigma_j^0)/2$}
\end{array}
\right.
\label{mm}
\end{equation}
and
\begin{equation}
U_{ij}^{10}(r)=  \left\{
\begin{array}{ll}
\infty\,, & \hbox{$r < (\sigma_i^1 + \sigma_j^0)/2$}\,,\\[5pt]
\displaystyle \frac{e^2z_i^1z_j^0}{4\pi \varepsilon_\mathrm{v} \varepsilon r}\,, &
\hbox{$r \geqslant (\sigma_i^1 + \sigma_j^0)/2$}\,,
\end{array}
\right.
\label{fm}
\end{equation}
\begin{equation}
U_{ij}^{11}(r) = \left\{
\begin{array}{ll}
\infty\,, &  \hbox{$r < (\sigma_i^1 + \sigma_j^1)/2$}\,,\\[5pt]
\displaystyle \frac{e^2z_i^1z_j^1}{4\pi \varepsilon_\mathrm{v} \varepsilon r}\,, &
\hbox{$r \geqslant (\sigma _i^1 + \sigma_j^1)/2$}\,.
\end{array}
\right.
\label{ff}
\end{equation}
In equations~(\ref{mm})--(\ref{ff}) $e$ denotes the elementary charge, $z_i^m$ $(z_j^m)$ is the nominal charge of ions $(m=0,1)$, $\varepsilon_\mathrm{v}$ is the permittivity of vacuum, $\varepsilon_0$ and $\varepsilon$ are the dielectric constants of the pre-quenching conditions and of the  partly-quenched system studied, respectively, $\sigma_i^0$ and $\sigma_i^1$ are the diameters of the matrix and of the annealed fluid particles, respectively, and as usual $r$ denotes the distance between particles $i$ and $j$. Note that in this work both matrix and annealed fluid are $+1:-1$ electrolytes ($|z_i^m| = 1$) with equal diameters, i.e., $\sigma_i^1=\sigma_i^0=4.25$~\AA.

The equilibrium structure of the matrix subsystem was obtained at temperature $T_0$ which is in general different from the temperature of observation, $T$. The relation between the two is given by the so-called quenching parameter $Q=\varepsilon_0 T_0/\varepsilon T$.

\section{Theoretical procedure \label{sec:theor}}

For systems containing Coulomb forces, a renormalization of the pair potential and correlation functions into short- and long-ranged terms is required to obtain numerical solution of the above given sets of integral equations \cite{ichiye1990,vlachy1991,duh1992,hribar1997,hribar1998}. While the renormalization scheme for the ROZ set of equations (\ref{roz}) is well established \cite{hribar1998,pqs11}, no such scheme for MGOZ equations (\ref{mgoz}) has been previously proposed. We follow the procedure described in detail in reference \cite{hribar1997} by splitting $h$ and $c$ functions into the short- and long-ranged parts
\begin{equation}
{\bf h}^{mn}={\bf h}_{(\mathrm{s})}^{mn}+{\bf q}^{mn}
\label{hsplit}
\end{equation}
and
\begin{equation}
{\bf c}^{mn}={\bf c}_{(\mathrm{s})}^{mn}+{{\boldsymbol \Phi }}^{mn}\,,
\label{hsplit2}
\end{equation}
where the superscripts $m,n$ assume the values $0,1$, and subscript (s)
denotes the short-ranged part of the correlation functions.
Since $+1:-1$ electrolyte is a two component system (cations and anions),
all the equations are written in the matrix form. Matrices are of the order
$2 \times 2$ and contain appropriate functions for $++$, $+-$, $-+$, and $--$
interactions. We choose the elements $\varphi_{ij}^{mn}(r)$ of the matrix
${\boldsymbol \Phi }^{mn}$ in the form of a Coulomb interaction \cite{hribar1997}
\begin{equation}
\begin{split}
\varphi _{ij}^{mn}(r) &=
-\frac{e^2z_i^mz_j^n}{4\pi \varepsilon_\mathrm{v} \varepsilon r k_\mathrm{B} T}\,,\\[5pt]
\varphi_{ij}^{00}(r)\ & =
-\frac{e^2z_i^0z_j^0}{4\pi \varepsilon_\mathrm{v} \varepsilon_0 r k_\mathrm{B} T_0}\,,
\end{split}
\label{phi}
\end{equation}
where $k_\mathrm{B}$ denotes the Boltzmann constant.

It is convenient to introduce the so-called Bjerrum length, $L_\mathrm{B} = e^2/(4 \pi\varepsilon_\mathrm{v} \varepsilon k_\mathrm{B} T)$, and the Debye-H\"{u}ckel screening lengths, $\kappa_0 = (4 \pi \rho^0 L_\mathrm{B}/Q)^{1/2}$ and $\kappa_1 = (4 \pi \rho^1 L_\mathrm{B})^{1/2}$. $\rho ^0 = \rho_{+} ^0 + \rho_{-} ^0$ and  $\rho^1 = \rho_{+} ^1 + \rho_{-} ^1$, where $\rho_{+}^m$ ($\rho_{-}^m$) is the number density of positive (negative) ions in the sub-system $m$. In the case of a symmetric electrolyte studied here $\rho_{+}^m = 0.5\rho^m = c_m N_\mathrm{A}$, where $c_m$ denotes the molar concentration of an electrolyte and $N_\mathrm{A}$ the Avogadro number.

By analogy with the procedure given in reference \cite{hribar1997}, we define the appropriate long-ranged total correlation functions, ${\bf q}$, through the equations
\begin{equation}
\begin{split}
{\bf q}^{10}-{\boldsymbol \Phi}^{10}={\boldsymbol \rho}^0{\boldsymbol \Phi}^{10}\otimes {\bf q}^{00} + {\boldsymbol \rho}^1{\boldsymbol \Phi}^{11}\otimes {\bf q}^{10},\\[5pt]
{\bf q}^{11}-{\boldsymbol \Phi}^{11}={\boldsymbol \rho}^0{\boldsymbol \Phi}^{10}\otimes {\bf q}^{10} + {\boldsymbol \rho}^1{\boldsymbol \Phi}^{11}\otimes {\bf q}^{11}.
\end{split}\label{q_def}
\end{equation}
The two equations (\ref{q_def}) can be readily solved to obtain the Fourier transforms of the screened potentials, and then inverted to the Cartesian space. For $+1:-1$ electrolyte, one obtains
\begin{equation}
\begin{split}
\begin{pmatrix}
q_{++}^{00}(r)&q_{+-}^{00}(r)\cr
q_{-+}^{00}(r)&q_{--}^{00}(r)\cr
\end{pmatrix}
&= - \frac{L_\mathrm{B}\exp (- \kappa_0 r)}{Qr}
\begin{pmatrix}
1 & -1 \cr
-1 & 1 \cr
\end{pmatrix}, \\[5pt]
\begin{pmatrix}
q_{++}^{10}(r) & q_{+-}^{10}(r) \cr
q_{-+}^{10}(r) & q_{--}^{10}(r) \cr
\end{pmatrix}
&= - L_\mathrm{B} \frac{\kappa_0^2}{\kappa_0^2 - \kappa_1^2}
 \left [ \frac{\exp (- \kappa_0 r)}{r} -
\frac{\kappa_1^2}{\kappa _0^2} \frac{\exp ( - \kappa_1 r)}{r}
\right ]
\begin{pmatrix}
1 & -1 \cr
-1 & 1 \cr
\end{pmatrix},  \\[5pt]
\begin{pmatrix}
q_{++}^{11}(r)&q_{+-}^{11}(r)\cr
q_{-+}^{11}(r)&q_{--}^{11}(r)\cr
\end{pmatrix}
&= -
\frac{L_\mathrm{B} \exp (- \kappa_1 r)}{r}
\begin{pmatrix}
1 & -1 \cr
-1 & 1 \cr
\end{pmatrix}
+  \frac{L_\mathrm{B} Q}{2}
\frac{\kappa_0^2 }{\kappa_0^2
- \kappa _1 ^2}   \\
&
\times\left [ -\frac{2 \kappa_0 ^2 \exp (-
\kappa _0 r)}{r (\kappa_0^2 - \kappa_1 ^2)} +
 \frac{ 2 \kappa_0^2 \exp (- \kappa_1 r)}{r (\kappa_0^2 - \kappa_1 ^2)}
- \kappa_1 \exp (- \kappa_1 r) \right ]
\begin{pmatrix}
1 & -1 \cr
-1 & 1 \cr
\end{pmatrix}.
\end{split}
\label{eq:qgunkcije1}
\end{equation}

It should be noted that in the case where $\kappa_0 = \kappa_1=\kappa$,
a somewhat different set of functions is obtained
\begin{equation}
\begin{split}
\begin{pmatrix}
q_{++}^{10}(r) & q_{+-}^{10}(r) \cr
q_{-+}^{10}(r) & q_{--}^{10}(r) \cr
\end{pmatrix}
&= - \left (1-\frac{\kappa}{2} r \right ) \frac{L_\mathrm{B} \exp (- \kappa r)}{r}
\begin{pmatrix}
1 & -1 \cr
-1 & 1 \cr
\end{pmatrix},
 \\[5pt]
\begin{pmatrix}
q_{++}^{11}(r)&q_{+-}^{11}(r)\cr
q_{-+}^{11}(r)&q_{--}^{11}(r)\cr
\end{pmatrix}
&= - \frac{L_\mathrm{B} \exp (- \kappa r)}{r} \left[ 1 - \frac{3}{8}
  \frac{\rho^0}{\rho^1} \kappa r \left (1 - \frac{1}{3} \kappa r \right ) \right ]
\begin{pmatrix}
1 & -1 \cr
-1 & 1 \cr
\end{pmatrix}.
\end{split}
\label{eq:qgunkcije2}
\end{equation}

By comparing the results (\ref{eq:qgunkcije1}) and (\ref{eq:qgunkcije2}) with the ${\bf q}$ functions obtained in work \cite{hribar1997} for the set of ROZ equations, one can see that the results are basically identical, suggesting that in the Debye-H\"{u}ckel limit both approaches provide the same results.

The renormalized MGOZ equations now read
\begin{equation}
\begin{split}
{\bf h}_{(\mathrm{s})}^{00}-{\bf c}_{(\mathrm{s})}^{00} &= {\bf c}_{(\mathrm{s})}^{00}\otimes
{\boldsymbol \rho}^0 \left ( {\bf h}_{(\mathrm{s})}^{00}+{\bf q}^{00} \right ) +
{\boldsymbol \Phi }^{00}\otimes {\boldsymbol \rho}^0{\bf h}_{(\mathrm{s})}^{00}\,, \\[5pt]
{\bf h}_{(\mathrm{s})}^{10}-{\bf c}_{(\mathrm{s})}^{10} &= {\bf c}_{(\mathrm{s})}^{10}
\otimes {\boldsymbol \rho}^0 \left ({\bf h}_{(\mathrm{s})}^{00}+{\bf q}^{00}
\right )+ {\boldsymbol \Phi }^{10}\otimes {\boldsymbol \rho}^0
{\bf h}_{(\mathrm{s})}^{00}+ {\bf c}_{(\mathrm{s})}^{11}\otimes {\boldsymbol \rho}^1
\left ( {\bf h}_{(\mathrm{s})}^{10} + {\bf q}^{10} \right ) +
{\boldsymbol \Phi }^{11}\otimes
{\boldsymbol \rho}^1{\bf h}_{(\mathrm{s})}^{10} \,,\\[5pt]
{\bf h}_{(\mathrm{s})}^{11}-{\bf c}_{(\mathrm{s})}^{11}&={\bf c}_{(\mathrm{s})}^{10}\otimes
{\boldsymbol \rho}^0\left ( {\bf h}_{(\mathrm{s})}^{01}+{\bf q}^{01}\right )+
{\boldsymbol \Phi }^{10}\otimes {\boldsymbol \rho}^0{\bf h}_{(\mathrm{s})}^{01}+
{\bf c}_{(\mathrm{s})}^{11}\otimes {\boldsymbol \rho}^1 \left (
{\bf h}_{(\mathrm{s})}^{11}+ {\bf q}^{11} \right )+
{\boldsymbol \Phi }^{11}\otimes
{\boldsymbol \rho}^1{\bf h}_{(\mathrm{s})}^{11}\,.
\end{split}
\end{equation}

To solve these equations, another relation between the total and direct correlation function is needed. Here, we used the hypernetted-chain approximation which reads~\cite{hribar1998}
\begin{equation}
c^{mn}(r)=\exp \left[-\beta U^{mn}(r)+\gamma^{mn}(r)\right]-1-\gamma^{mn}(r),
\label{hnc}
\end{equation}
where $\gamma^{mn}(r)=h^{mn}(r)-c^{mn}(r)$ and the superscripts $m,n$ assume the values $0$ and $1$. $U^{mn}(r)$ are the inter-particle potentials for different components [equations (\ref{mm})--(\ref{ff})] and $\beta = 1/k_\mathrm{B}T$.

MGOZ equations were solved by a direct iteration on a grid of $2^{15}$ points with 0.005~{\AA} spacing in $r$-space to obtain appropriate correlation functions (the same was done for ROZ equations). These were then used to calculate various thermodynamic properties using the well-established equations \cite{kierlik1997,rosinberg1994,belloni1988,hribar1998,hribar2000} adapted for the MGOZ approximation. The reduced excess internal energy per an annealed fluid particle can be calculated via the expression
\begin{equation}
\frac{\beta E}{N_1} = \frac{1}{2} \sum_{i=+,-} \sum_{j=+,-}
x_i^{1} \rho_j^1 \int g_{ij}^{11}(r)U_{ij}^{11}(r) \mathrm{d}{\bf r} +\sum_{i=+,-} \sum_{j=+,-} x_i^{1}
\rho_j^0 \int g_{ij}^{10}(r)U_{ij}^{10}(r) \mathrm{d}{\bf r}.
 \label{15}
\end{equation}
The isothermal compressibility reads
\begin{equation}
\left (\frac{\partial \beta P}{\partial \rho^1} \right )_T = 1- \rho^1 \sum_{i=+,-} \sum_{i=+,-} x_i^{1} x_j^{1} \int c_{(\mathrm{s})ij}^{11}(r) \mathrm{d}{\bf r}.
\label{eq:compress}
\end{equation}
The excess chemical potential of species $i$ (logarithm of the activity coefficient $\gamma_i$) is
\begin{eqnarray}
\hspace{-4mm}\beta \mu_{i,1}\hspace{-2mm} &=& \hspace{-2mm} -{ \sum_{j=+,-}} \rho_j^0 \int c_{({s})ij}^{10}\rd{\bf r} -{ \sum_{j=+,-}} \rho _j^1 \int c_{({s})ij}^{11}\rd{\bf r} \nonumber\\
&&+ \frac{1}{2}{ \sum_{j=+,-} }\rho_j ^0 \int h_{ij}^{10}(r) \left [
h_{ij}^{10}(r)-c_{ij}^{10}(r) \right ] \mathrm{d}{\bf r} \,\,\,+ 
\frac{1}{2}{ \sum_{j=+,-} } \rho_j^1\int
\left \{ h_{ij}^{11}(r) \left [ h_{ij}^{11}(r)-c_{ij}^{11}(r)\right ]
\right \}
\mathrm{d}{\bf r}.
\label{beloni-eq}
\end{eqnarray}
In the above equations, $x_i^1 = \rho_i^1/\sum_i \rho_i^1$ denotes the mole fraction of the annealed species, $g_{ij}^{mn}(r) = h_{ij}^{mn}(r) + 1$ is the radial distribution function, $U_{ij}^{mn}(r)$ is the pair potential ($m,n=0,1$), and $\mathrm{d}{\bf r} = 4\pi r^2 \mathrm{d}r$. Note that due to MG approximations, equations (\ref{eq:compress}) and (\ref{beloni-eq}) are somewhat different form those given in references \cite{kierlik1997,rosinberg1994,belloni1988,hribar1998,hribar2000} for ROZ approach.

\section{Results}

\begin{figure}[!t]
\centerline{
\includegraphics[width=1.0\textwidth]{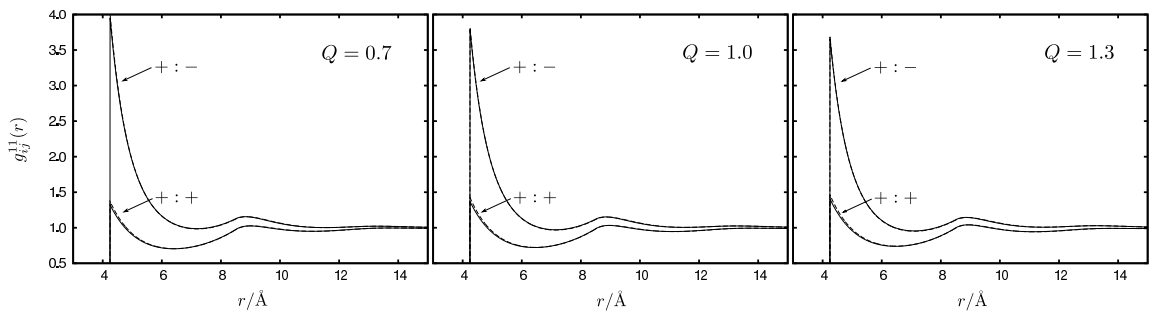}
}
\caption{Radial distribution functions of the annealed electrolyte, $g^{11}_{ij}(r)$, for the case $c_1 = 1.0$~mol dm$^{-3}$ and $c_0 = 5.0$~mol dm$^{-3}$. Quenching parameters were $Q = 0.7$, $1.0$, and $1.3$. Solid lines show the ROZ and dashed lines show the MGOZ approach.}
\label{fig:gr11}
\end{figure}
\begin{figure}[!b]
\centerline{
\includegraphics[width=1.0\textwidth]{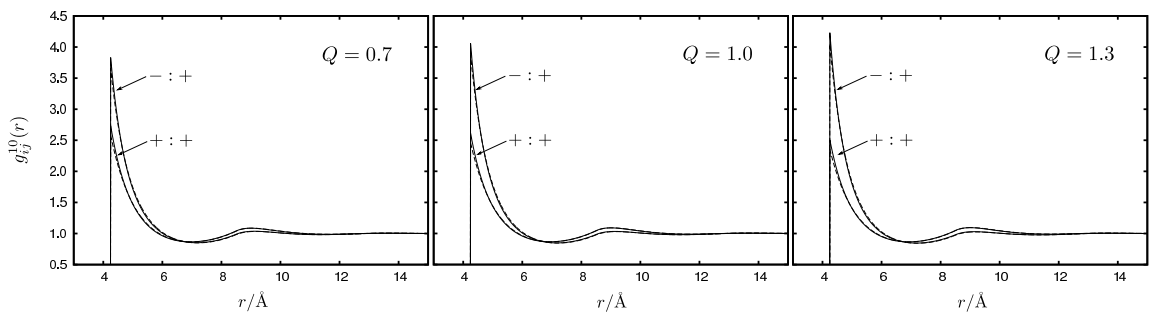}
}
\caption{The same as in figure~\ref{fig:gr11} but for the interaction of the annealed particles with the matrix particles, $g^{10}_{ij}(r)$. In the notation $-:+$, the minus sign applies to the matrix and plus to the annealed ions.}
\label{fig:gr10}
\end{figure}

We show the results for the radial distribution functions and thermodynamic quantities (reduced excess internal energy, excess chemical potential, and isothermal compressibility) for selected examples of the partly-quenched system studied. In all the cases, the temperature of observation was taken to be $T=298$~K and the dielectric constant of aqueous solutions at this $T$ was $\varepsilon = 78.5$. The Bjerrum length equals $L_\mathrm{B} = 7.14$~{\AA} for the chosen $T$ and $\varepsilon$.

As already explained in the above section, the ${\bf q}^{10}$ and ${\bf q}^{11}$ functions for the case of MGOZ approach are the same as those presented in reference \cite{hribar1997} for ROZ equations. We, therefore, examined the total correlation functions in the form of pair distribution functions, $g^{mn}_{ij}(r)$. These are for three different quenching parameters ($Q= 0.7$, 1.0, and 1.3), and one matrix concentration ($c_0 = 5.0$~mol dm$^{-3}$) presented in figures~\ref{fig:gr11} and \ref{fig:gr10}. In both cases the annealed fluid concentration was $c_1 = 1.0$~mol dm$^{-3}$. The results for the ROZ equations are shown by solid lines and those of MGOZ equations by dashed lines.

One can see that there is no obvious difference between the two results. This is expected since the behaviour of the  systems studied was found to be mainly determined by the long-ranged correlations \cite{hribar1998}.

\begin{figure}[!t]
\centerline{
\includegraphics[width=1.0\textwidth]{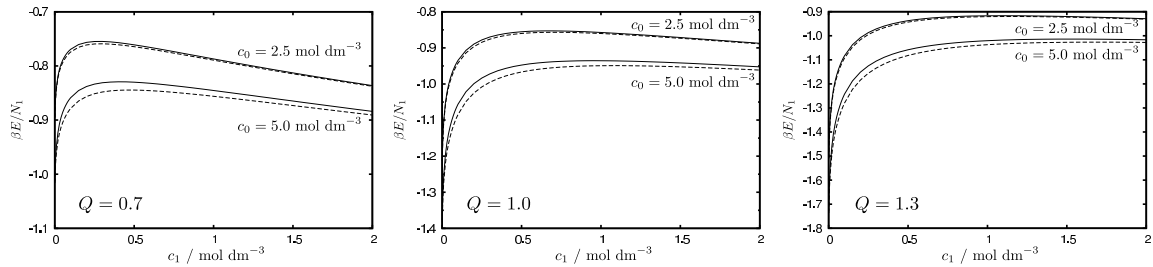}
}
\caption{The reduced excess internal energy per annealed particle as a function of the annealed electrolyte concentration. Matrix concentrations were 2.5 and 5.0~mol dm$^{-3}$, and quenching parameters $Q = 0.7$, $1.0$, and $1.3$. Solid lines apply to the Give-Stell and dashed to Madden-Glandt equations.}
\label{fig:energy}
\end{figure}

\begin{figure}[!h]
\centerline{
\includegraphics[width=1.0\textwidth]{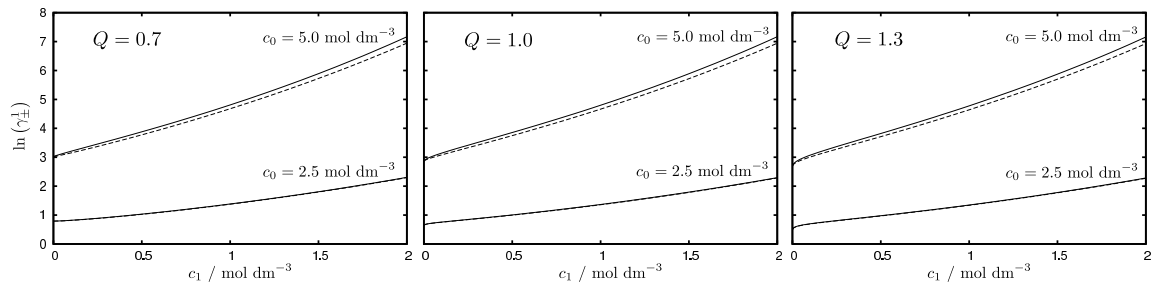}
}
\caption{The logarithm of the mean activity coefficient, i.e., the excess chemical potential, of the annealed electrolyte. Parameters and notations are the same as in figure~\ref{fig:energy}.}
\label{fig:gamma}
\end{figure}

\begin{figure}[!h]
\centerline{
\includegraphics[width=1.0\textwidth]{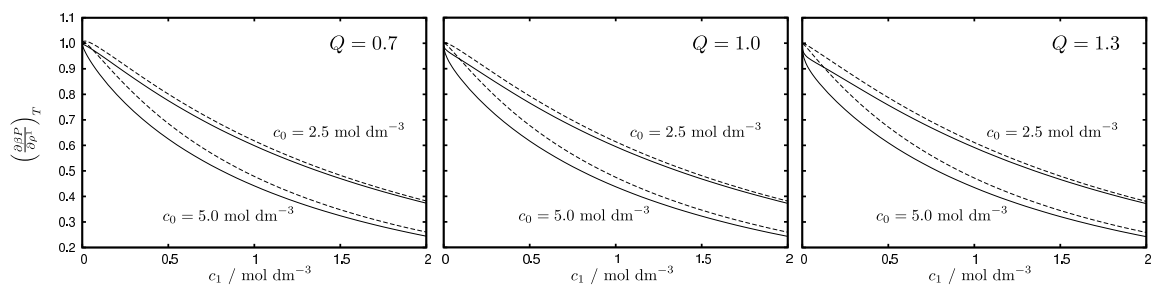}
}
\caption{The isothermal compressibility of the annealed electrolyte. Parameters and notations are the same as in figure~\ref{fig:energy}.}
\label{fig:compress}
\end{figure}

We proceed with the results for thermodynamic properties, namely the reduced excess internal energy, the excess chemical potential, and the isothermal compressibility. The results for different conditions  studied are presented in figures~\ref{fig:energy}--\ref{fig:compress}.

As previously observed in work \cite{luksic2007}, the excess internal energy (at a given matrix concentration) first increases with the increasing concentration of the annealed electrolyte, and then slowly starts to decrease. This is a consequence of the annealed fluid interaction with the matrix particles. Both approaches, ROZ (solid lines) and MGOZ (dashed lines), provide very similar results. The differences, as expected, become larger at higher matrix concentrations.

Figure~\ref{fig:gamma} shows the results for the excess chemical
potential, $\beta \mu_1 = \ln \gamma_{\pm}^1$, as a function of the annealed electrolyte concentration within the two matrices with different concentrations. Similarly to the excess internal energy, very small differences are observed between the MGOZ and ROZ results. The MGOZ lie slightly below the ROZ results for high matrix and annealed fluid concentrations.

Small but more pronounced  differences between the two approaches are still observed for  isothermal compressibility. These differences pertain even at low annealed fluid concentrations, where short-ranged interactions become important.

\section{Conclusions}
A renormalization scheme for the MGOZ equations was proposed and used to study the structural and thermodynamic properties of simple systems, where the quenched and annealed subsystems were both $+1:-1$ restricted primitive model electrolytes within the hypernetted-chain approximation. The results were compared with those obtained by ROZ/HNC approximation.

The results obtained for the ${\bf q}$ functions determining the long-ranged behaviour of the total correlation functions are identical to those evaluated for the ROZ equations. Consequently, no significant differences in the structural properties were observed. The differences between the two approaches become more important when studying the compressibility of the system, where short-ranged interactions play a central role.

\section*{Acknowledgements}

Financial support of the Slovenian Research Agency (ARRS) through grant P1-0201 and project J1-4148 is acknowledged. M. L. also acknowledges
the funding through the Research Foundation of SUNY.

\newpage

\ukrainianpart

\title
{Структура і термодинаміка примітивної моделі розчину електроліту у
зарядженій матриці: \\ Дослідження наближення Маддена-Гландта }
\author{М. Лукшіч\refaddr{label1,label2}, Б. Грібар-Лі\refaddr{label2}}
\addresses{
\addr{label1} Університет Любляни, Факультет хімії і хімічної
технології,  SI-1000 Любляна, Словенія \addr{label2} Державний
 Нью-Йоркський  університет в Стоні Брук, Стоні Брук, NY 11794-5252, США}

\makeukrtitle

\begin{abstract}
\tolerance=3000%
Ми порівняли результати наближення Маддена-Гландта у формалізмі
інтегральних рівнянь для частково заморожених систем із загально
прийнятим формулюванням Гівена і Стелла. Розглядається  $+1:-1$
валентна обмежена примітивна модель розчину електроліту у
замороженій $+1:-1$ валентній матриці. Запропоновано схему
перенормування для наближення Маддена-Гландта реплікованого рівняння
Орнштейна-Церніке. Виявляється, що далекосяжні вклади до прямої і
повної кореляційних функцій, які описують взаємодію між частинками
електроліту у певній репліці, а також між частинками плину і матриці,
є однаковими у наближенні Маддена-Гландта і формулюванні Гівена і
Стелла. Обидві теорії призводять до дуже подібних результатів для
структури і термодинаміки підсистеми адсорбату. Різниця між
значеннями надлишкової внутрішньої енергії, надлишкового хімічного
потенціалу та ізотермічної стисливості стає більш виразною лише при
високій густині матричної підсистеми.
\keywords частково заморожені системи, розчини електролітів,
рівняння Орнштейна-Церніке, теорія реплік, структура, термодинаміка

\end{abstract}


\begin{thebibliography}{99}

\bibitem{pqs11} Hribar-Lee~B., Luk\v{s}i\v{c}~M., Vlachy~V.,
  Annu. Rep. Prog. Chem., Sect. C, 2011,
    \textbf{107}, 14; \doi{10.1039/c1pc90001c}.

\bibitem{given1992} Given~J.A., Stell~G.,
 J. Chem. Phys., 1992,     \textbf{97}, 4573; \doi{10.1063/1.463883}.

\bibitem{given1994} Given~J.A., Stell~G.,
 Physica A, 1994,
    \textbf{209}, 495; \doi{10.1016/0378-4371(94)90200-3}.

\bibitem{pizio1998} Pizio~O., Sokolowski~S.,
J. Phys. Stud., 1998,  \textbf{2}, 296.

\bibitem{rosinberg1999} Rosinberg~M.L., In: New Approaches to Problems
  in Liquid State Theory, Caccamo~C., Hansen~J.P., Stell~G. (Eds.),
  Kluwer, Dordrecht, 1999, 245--278.

\bibitem{pizio2000} Pizio~O., In: Computational Methods in Surface and Colloid Science,
Surfactant science series, Vol.~89, Borowko~M. (Ed.),
  Kluwer, Marcel Dekker, New York, 2000, 293--345.

\bibitem{trokhymchuk1996} Trokhymchuk~A.D., Pizio~O.,  Holovko~M.F,  Sokolowski~S.,
 J. Phys. Chem., 1996,
    \textbf{100}, 17004; \doi{10.1021/jp961443l}.

\bibitem{madden1988} Madden~W.G., Glandt~E.D.,
 J. Stat. Phys., 1988,
    \textbf{51}, 537; \doi{10.1007/BF01028471}.

\bibitem{madden1992} Madden~W.G.,
 J. Chem. Phys., 1992,     \textbf{96}, 5422; \doi{10.1063/1.462726}.

\bibitem{ichiye1990} Ichiye~T., Haymet~A.D.J.,
 J. Chem. Phys., 1990,
    \textbf{93}, 8954; \doi{10.1063/1.459234}.

\bibitem{duh1992} Duh~D.M., Haymet~A.D.J.,
 J. Chem. Phys., 1992,
    \textbf{97}, 7716; \doi{10.1063/1.463491}.

\bibitem{vlachy1991} Vlachy~V., Ichiye~T., Haymet~A.D.J.,
 J. Am. Chem. Soc., 1991,
    \textbf{113}, 1077; \doi{10.1021/ja00004a003}.

\bibitem{hribar1997} Hribar~B., Pizio~O., Trokhymchuk~A., Vlachy~V.,
 J. Chem. Phys., 1997,
    \textbf{107}, 6335; \doi{10.1063/1.474294}.

\bibitem{hribar1998} Hribar~B., Pizio~O., Trokhymchuk~A., Vlachy~V.,
 J. Chem. Phys., 1998,
    \textbf{109}, 2480; \doi{10.1063/1.476819}.

\bibitem{bratko1996} Bratko~D., Chakraborty~A.K.,
 J. Chem. Phys., 1996,
    \textbf{104}, 7700; \doi{10.1063/1.471476}.

\bibitem{hribar2000} Hribar~B., Vlachy~V., Pizio~O.,
 J. Phys. Chem. B, 2000,
    \textbf{104}, 4479; \doi{10.1021/jp994324p}.

\bibitem{kierlik1997} Kierlik~E., Rosinberg~M.L., Tarjus~G., Monson~P.A.,
 J. Chem. Phys., 1997,
    \textbf{106}, 264; \doi{10.1063/1.474134}.

\bibitem{rosinberg1994} Rosinberg~M.L., Tarjus~G., Stell~G.,
 J. Chem. Phys., 1994,
    \textbf{100}, 5172; \doi{10.1063/1.467182}.

\bibitem{belloni1988} Belloni~L.,
 J. Chem. Phys., 1988,
    \textbf{88}, 5143; \doi{10.1063/1.454668}.

\bibitem{luksic2007} Luk\v{s}i\v{c}~M., Hribar-Lee~B., Vlachy~V.,
 Acta Chim. Slov., 2007,
    \textbf{54}, 523.

\end{thebibliography}
\end{document}